\documentclass[a4paper,aps,prd,10pt,preprintnumbers,showpacs,showkeys,twocolumn,superscriptaddress,nofootinbib,amsmath,amssymb]{revtex4-1}
\usepackage{cmap}
\usepackage[utf8]{inputenc}
\usepackage[T1]{fontenc}
\usepackage{graphicx,orcidlink,xcolor,ulem}
\usepackage{bm}

\begin{document}

\title{Telling tails and quasi-resonances in the vicinity of Dymnikova regular black hole}

\author{Bekir Can Lütfüoğlu}
\email{bekir.lutfuoglu@uhk.cz}
\affiliation{Department of Physics, Faculty of Science, University of Hradec Králové, Rokitanského 62/26, 500 03 Hradec Králové, Czech Republic}

\author{Javlon Rayimbaev}
\email{javlon@astrin.uz}
\affiliation{Institute of Theoretical Physics, National University of Uzbekistan, Tashkent 100174, Uzbekistan}
\affiliation{University of Tashkent for Applied Sciences, Str. Gavhar 1, Tashkent 100149, Uzbekistan} \affiliation{Tashkent State Technical University, Tashkent 100095, Uzbekistan}


\author{Bekzod~Rahmatov}
\email{rahmatovbekzod@samdu.uz}
\affiliation{New Uzbekistan University, Movarounnahr Street 1, Tashkent 100007, Uzbekistan}

\author{Fayzullo~Shayimov}
\email{shayimovf@gmail.com}
\affiliation{Kimyo International University in Tashkent, Shota Rustaveli street 156, Tashkent 100121, Uzbekistan}


\author{Ikram Davletov}
\email{ikram.d@urdu.uz}
\affiliation{Department of Technique, Urgench State University, Kh. Alimjan Str. 14, Urgench 221100, Uzbekistan}



\begin{abstract}
We investigate quasinormal modes, late-time tails, and grey-body factors for massive scalar perturbations in the background of the Dymnikova regular black hole. By applying both the time-domain integration and the WKB method with Padé improvements, we show that the spectrum of massive fields differs qualitatively from the massless case. The oscillation frequency of the dominant mode grows with the field mass $\mu$, while the damping rate decreases, suggesting the existence of quasi-resonances at sufficiently large $\mu$. In the time domain, the late-time signal exhibits oscillatory tails with a power-law envelope, whose decay rate matches analytic expectations. Grey-body factors are also computed, showing strong suppression of radiation when mass is increased. Taken together, these results indicate that massive fields provide distinctive signatures of regular black holes and may serve as probes of near-horizon quantum corrections in the Dymnikova geometry.
\end{abstract}

\maketitle

\section{Introduction}

The study of quasinormal modes (QNMs) \cite{Kokkotas:1999bd,Berti:2009kk,Konoplya:2011qq,Bolokhov:2025uxz} has become one of the central tools in black-hole physics, providing a bridge between theory and observation in the era of gravitational-wave astronomy \cite{LIGOScientific:2016aoc, LIGOScientific:2017vwq, LIGOScientific:2020zkf,KAGRA:2013rdx}. While the bulk of the literature has focused on massless perturbations, massive fields bring in a number of qualitatively new phenomena that significantly enrich the spectral structure of black holes. Quasinormal oscillations of massive fields of different spins have been analyzed in a wide variety of settings (see, e.g., \cite{Konoplya:2004wg,Konoplya:2013rxa,Konoplya:2017tvu,Zhidenko:2006rs,Ohashi:2004wr,Zhang:2018jgj,Aragon:2020teq,Ponglertsakul:2020ufm,Gonzalez:2022upu,Burikham:2017gdm,Skvortsova:2025cah} and references therein). These studies have shown that the inclusion of a mass parameter for the field gives rise to several intriguing effects. For instance, the effective mass term may naturally arise in higher-dimensional braneworld scenarios due to the influence of the bulk on brane-localized fields \cite{Seahra:2004fg}. In addition, massive gravitons in modified gravity frameworks have been invoked as possible contributors to ultra-long wavelength signals currently probed by Pulsar Timing Array experiments \cite{Konoplya:2023fmh,NANOGrav:2023hvm}.  

Another remarkable phenomenon is the existence of arbitrarily long-lived oscillations at specific values of the field’s mass, often referred to as quasi-resonances \cite{Konoplya:2004wg,Ohashi:2004wr}. This effect is rather general, spanning different field spins \cite{Fernandes:2021qvr,Konoplya:2017tvu,Percival:2020skc}, a variety of black-hole backgrounds \cite{Konoplya:2013rxa,Zhidenko:2006rs,Zinhailo:2018ska,Churilova:2020bql,Bolokhov:2023bwm}, and even exotic compact objects such as wormholes \cite{Churilova:2019qph}. At the same time, quasi-resonances are not a universal feature: there are known counterexamples in which massive perturbations do not give rise to infinitely long-lived modes \cite{Zinhailo:2024jzt}. Beyond the resonance structure, the late-time tails of massive perturbations also differ qualitatively from the power-law decay of massless fields, instead exhibiting oscillatory behavior, a phenomenon studied in detail in \cite{Jing:2004zb,Koyama:2001qw,Moderski:2001tk,Rogatko:2007zz,Koyama:2001ee,Koyama:2000hj,Gibbons:2008gg,Gibbons:2008rs,Lutfuoglu:2025hwh}. Moreover, even a massless field propagating in the presence of external fields, such as a magnetic background, effectively acquires a mass term in its perturbation equations \cite{Konoplya:2008hj,Wu:2015fwa,Kokkotas:2010zd}, further motivating the exploration of massive cases.  

In parallel, increasing attention has been devoted to black-hole geometries free of curvature singularities, motivated by both quantum-gravity considerations and astrophysical modeling. Among the earliest and most studied models is the Dymnikova spacetime \cite{Dymnikova:1992ux}, where the central singularity is replaced by a de Sitter core, resulting in a regular black hole. More recently, the same functional form has been shown to arise naturally within the framework of Asymptotic Safety, through renormalization-group improvements of the Schwarzschild solution \cite{Platania:2019kyx}. Thus, the Dymnikova black hole provides a useful testing ground for probing the interplay between regularity, quantum corrections, and perturbation dynamics.  

Quasinormal modes of {\it massless} test fields in the Dymnikova geometry have been analyzed in a number of recent works \cite{Konoplya:2024kih,Macedo:2024dqb,Konoplya:2023aph,Dubinsky:2025nxv}, while gravitational perturbations and quasinormal spectrum was considered in \cite{Dymnikova:2004qg,Lutfuoglu:2025pzi}, establishing the basic stability properties and spectral shifts induced by the de Sitter core. However, to the best of our knowledge, the spectrum of {\it  massive} fields in this background has not yet been addressed. Given the rich phenomenology associated with massive perturbations, ranging from quasi-resonances to oscillatory tails, and their potential observational consequences, it is both natural and timely to extend the analysis to this case. The present paper fills this gap by studying QNMs of a massive scalar field in the background of the Dymnikova black hole, thereby providing new insights into the dynamics of regular black holes inspired by quantum-gravity corrections.

The body of work on QNMs of quantum-corrected black holes is already extensive. Here we limit ourselves to pointing out only a representative selection of recent contributions~\cite{Bolokhov:2025lnt,Chen:2023wkq,Bonanno:2025dry,Malik:2024elk,Fu:2023drp,Abdalla:2005hu,Baruah:2023rhd,Konoplya:2020der,Malik:2025dxn,Skvortsova:2023zca,Heidari:2023ssx,Bolokhov:2023ozp,Kokkotas:2017zwt,Skvortsova:2024msa,Moreira:2023cxy,Konoplya:2017lhs}, and we refer the interested reader to the citations within those works for a broader perspective.

The paper is organized in the following way. In Sec. II, we introduce Dymnikova geometry for the regular black hole. Section III is devoted to perturbation equations and boundary conditions for a massive scalar field. In Sec. IV, we study quasinormal modes, asymptotic tails, and discuss the relation of quasinormal modes with grey-body factors. Finally, in the Conclusions, we summarize the obtained results.

\section{The Dymnikova Black Hole Geometry}\label{sec:dymnikova}

The requirement of regularity is of central importance in any consistent theory of gravity, as physical spacetimes are expected to be free of curvature singularities that signal the breakdown of classical general relativity. Regular black-hole solutions thus provide valuable laboratories for exploring how quantum or effective corrections may cure such pathologies while preserving key macroscopic features of black holes. The search for black-hole spacetimes free of curvature singularities has led to a variety of regular models in which the central divergence of the Schwarzschild solution is removed. One of the earliest and most widely cited examples is the construction of Dymnikova~\cite{Dymnikova:1992ux}, which replaces the central singularity with a smooth de Sitter core, while preserving asymptotic flatness at large distances. This requires a stress--energy tensor that technically violates certain classical energy conditions, but can be regarded as an effective description of quantum backreaction. The resulting geometry is static, spherically symmetric, everywhere regular, and asymptotically flat, with all curvature scalars finite.  

The spacetime metric keeps the familiar Schwarzschild structure,
\begin{equation}\label{metric}
ds^{2} = - f(r)\, dt^{2} + \frac{dr^{2}}{f(r)} + r^{2}\,(d\theta^{2} + \sin^{2}\theta\, d\phi^{2}),
\end{equation}
but with the lapse function determined by a nontrivial mass profile,
\begin{equation}
f(r) = 1 - \frac{2 M(r)}{r}.
\end{equation}
Instead of a constant mass, Dymnikova suggested \cite{Dymnikova:2003vt,Dymnikova:1992ux}
\begin{equation}
M(r) = M\left(1 - e^{-r^{3}/r_{0}^{3}}\right),
\end{equation}
where $M$ is the ADM mass and $r_{0}$ defines the de Sitter core radius. At short distances one finds
\[
f(r) \simeq 1 - \frac{r^{2}}{r_{0}^{2}},
\]
mimicking a regular de Sitter interior, while for $r\gg r_{0}$ the usual Schwarzschild form $f(r)\simeq 1 - 2M/r$ is recovered. The solution thus interpolates smoothly between the de Sitter and Schwarzschild limits.  

\medskip

Interestingly, the same functional structure was later rediscovered in the context of Asymptotic Safety~\cite{Platania:2019kyx}. Starting from the classical lapse,
\begin{equation}\label{lapse}
f(r) = 1 - \frac{2M}{r},
\end{equation}
the idea is to replace Newton’s constant by a scale-dependent running coupling. A convenient parametrization is
\begin{equation}\label{runningG}
G(r) = \frac{G_{0}}{1 + \tfrac{G_{0}}{g_{\ast}}\, k^{2}(r)},
\end{equation}
where $G_{0}$ is the low-energy value, $g_{\ast}$ the nontrivial fixed point, and $k(r)$ an effective cutoff that vanishes at infinity so that the Schwarzschild solution is restored asymptotically. Inserting this into the metric leads to
\begin{equation}
f(r) = 1 - \frac{2M}{r}\,\frac{G(r)}{G_{0}}.
\end{equation}

The modified geometry can be interpreted as if it were supported by an effective anisotropic fluid, with stress tensor
\begin{equation}\label{Teff}
T_{\mu\nu}^{\rm eff} = (\rho + p)(l_{\mu}n_{\nu}+l_{\nu}n_{\mu}) + p g_{\mu\nu},
\end{equation}
where $l_{\mu} n^{\mu} = -1$. The associated effective pressure and energy density follow from the radial variation of $G(r)$,
\begin{equation}\label{rho-p}
\rho = \frac{M G'(r)}{4\pi r^{2} G(r)}, 
\qquad 
p = -\frac{M G''(r)}{8\pi r\, G(r)}.
\end{equation}
This can be interpreted as the imprint of quantum vacuum polarization, since departures of $G(r)$ from constancy act as effective sources that modify the geometry.  

\medskip

By iteratively updating the cutoff $k(r)$ as a functional of the induced energy density, one obtains a self-consistent renormalization-group improvement of the metric. In the continuum limit the process converges to the explicit expression
\begin{equation}\label{dymnikova-final}
f(r) = 1 - \frac{2M}{r}\left(1 - e^{-r^{3}/(2 l_{\rm cr}^{2} M)}\right),
\end{equation}
which exactly coincides with the Dymnikova metric. Here $l_{\rm cr}$ is a critical length characterizing the scale of quantum corrections: for $l_{\rm cr}=0$ the Schwarzschild geometry is recovered, while finite $l_{\rm cr}$ generates a regular de Sitter-like core. Horizons exist only when
\[
l_{\rm cr} \lesssim 1.138\,M,
\]
beyond which the configuration becomes a horizonless, nonsingular compact object. In what follows, we adopt units $M=1$, so that all quantities are expressed in dimensionless form.  

The event horizon radius $r_h$ is determined implicitly by the condition
\begin{equation}\label{horizon-eq}
r_h - 2M \left(1 - e^{-\tfrac{r_h^{3}}{2 l^{2} M}}\right) = 0.
\end{equation}
The Hawking temperature is then given by
\begin{equation}\label{TH}
T_{H} = \frac{1}{4\pi}\left[\frac{1}{r_h} + \frac{3r_h}{l^{2}}
\left(1 - \frac{r_h}{2M}\right)\right].
\end{equation}

\medskip

The Dymnikova black hole thus enjoys a dual interpretation: historically, it was a phenomenological attempt to resolve singularities within classical general relativity, while modern renormalization-group arguments show it to arise naturally from quantum corrections in Asymptotic Safety. This makes it an especially compelling model of a regular black hole. Moreover, higher-dimensional generalizations have been identified in theories with curvature-squared corrections~\cite{Konoplya:2024kih}, underlining its role as a versatile testing ground for semiclassical and quantum-gravity effects.

\section{Massive scalar-field perturbations}
\label{sec:massive_scalar}

\begin{figure*}
\resizebox{\linewidth}{!}{\includegraphics{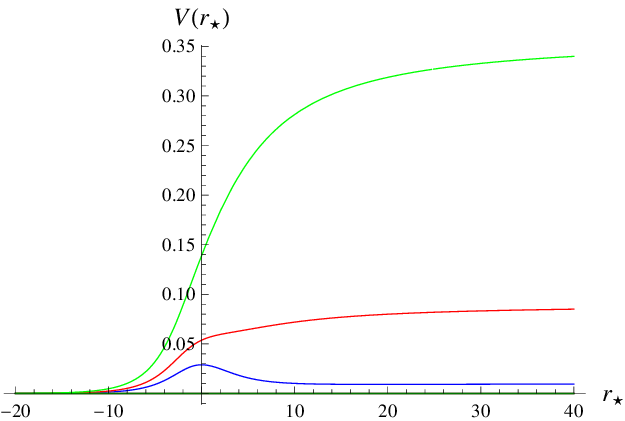}\includegraphics{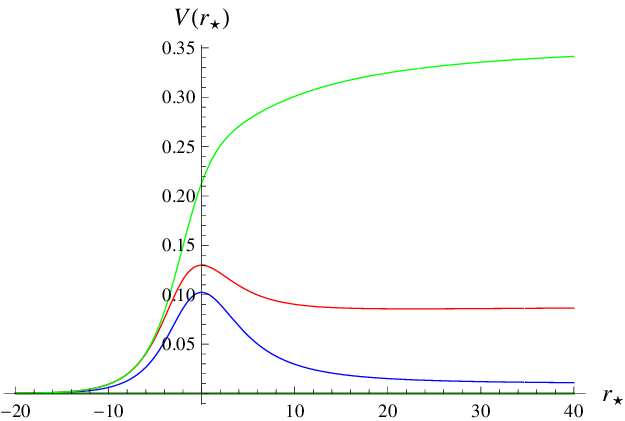}}
\caption{Effective potential as a function of the tortoise coordinate $r^{*}$ for $\ell=0$ (left) and $\ell=1$ (right), $\mu=0.1$ (blue), $\mu=0.3$ (red), $\mu=0.6$ (green); $l_{cr}=1.137$.}\label{fig:gravpot1}
\end{figure*}


In this section we consider a minimally coupled massive scalar field of mass $\mu$ propagating on the Dymnikova background \eqref{metric}. The field obeys the Klein--Gordon equation,
\begin{equation}
\Box \Phi - \mu^{2}\,\Phi = 0,
\qquad
\Box \equiv g^{\mu\nu}\nabla_{\mu}\nabla_{\nu}.
\label{eq:KG}
\end{equation}
Using the standard separation ansatz
\begin{equation}
\Phi(t,r,\theta,\phi)=\frac{1}{r}\,\Psi(r)\,Y_{\ell m}(\theta,\phi)\,e^{-i\omega t},
\label{eq:ansatz}
\end{equation}
and introducing the tortoise coordinate via $dr_{*}/dr = 1/f(r)$, one reduces \eqref{eq:KG} to the Schr\"odinger-like master equation
\begin{equation}
\frac{d^{2}\Psi}{dr_{*}^{2}}+\Bigl(\omega^{2}-V_{\ell}(r)\Bigr)\Psi=0,
\label{eq:RadialEq}
\end{equation}
with the effective potential
\begin{equation}
V_{\ell}(r)=f(r)\left[\mu^{2}+\frac{\ell(\ell+1)}{r^{2}}+\frac{f'(r)}{r}\right],
\quad \ell=0,1,2,\ldots
\label{eq:Vscalar}
\end{equation}
where $f'(r)\equiv df/dr$. 

The boundary conditions for quasinormal frequencies imply the purely outgoing wave at infinity and purely incoming on at the event horizon 
\begin{equation}
\Psi(r) \propto
\begin{cases}
e^{- i \omega r_{*}}, & r \to r_{+} ,\\[4pt] e^{+ i \Omega r_{*}}, & r \to \infty,
\end{cases}
\label{eq:BCs}
\end{equation}
where we used
\begin{equation}
\Omega = \sqrt{\omega^{2} - \mu^{2}},
\end{equation}
and the square root is fixed in such a way that $\mathrm{Re}(\Omega)$ and $\mathrm{Re}(\omega)$ are of the same sign \cite{Zhidenko:2006rs}.

The above boundary conditions select a discrete set of complex frequencies $\omega=\omega_{R}-i\,\omega_{I}$ ($\omega_{I}>0$). For certain ranges of parameters $(\mu,\ell)$, the imaginary part can become arbitrarily small (``quasi-resonances''), while the late-time signal at fixed radius exhibits oscillatory massive tails rather than the massless power-law decay.

\begin{figure*}
\centering
\includegraphics[width=0.48\linewidth]{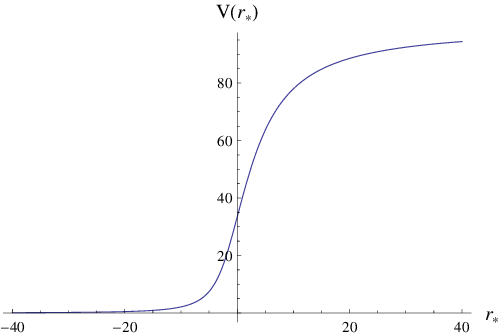}
\includegraphics[width=0.48\linewidth]{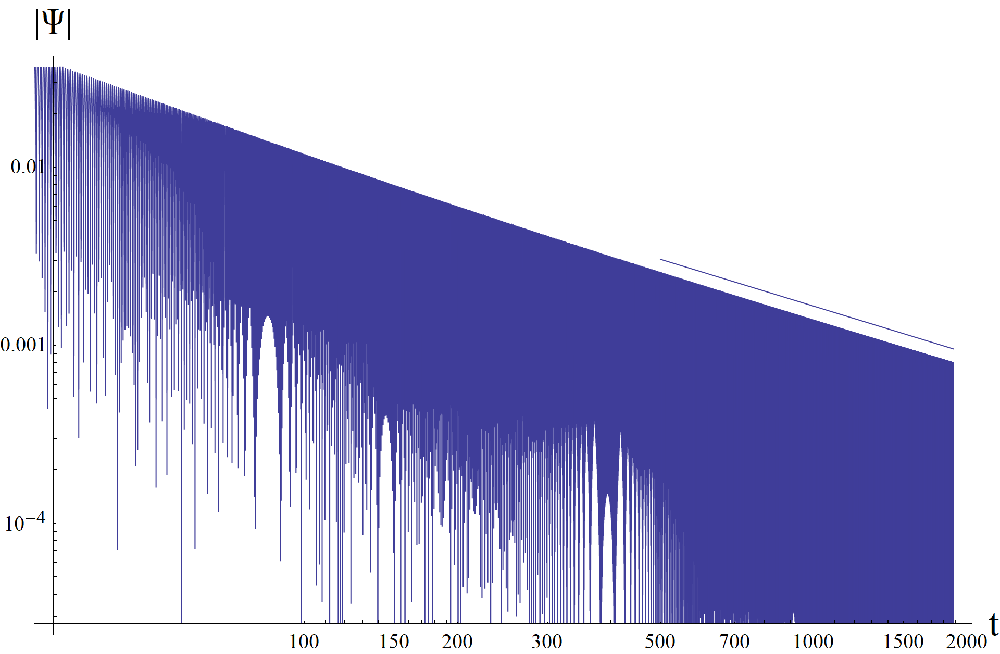}
\caption{Effective potential and logarithmic time-domain profile together with the line $\sim t^{-7/8}$ for $\ell=0$, $\mu=10$, $l_{cr}=1.137$, $M=1$.}\label{fig:L0}
\end{figure*}

\begin{figure*}
\centering
\includegraphics[width=0.48\linewidth]{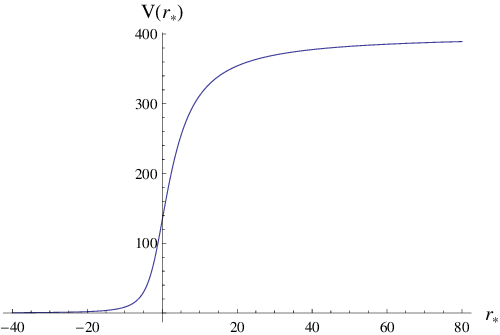}
\includegraphics[width=0.48\linewidth]{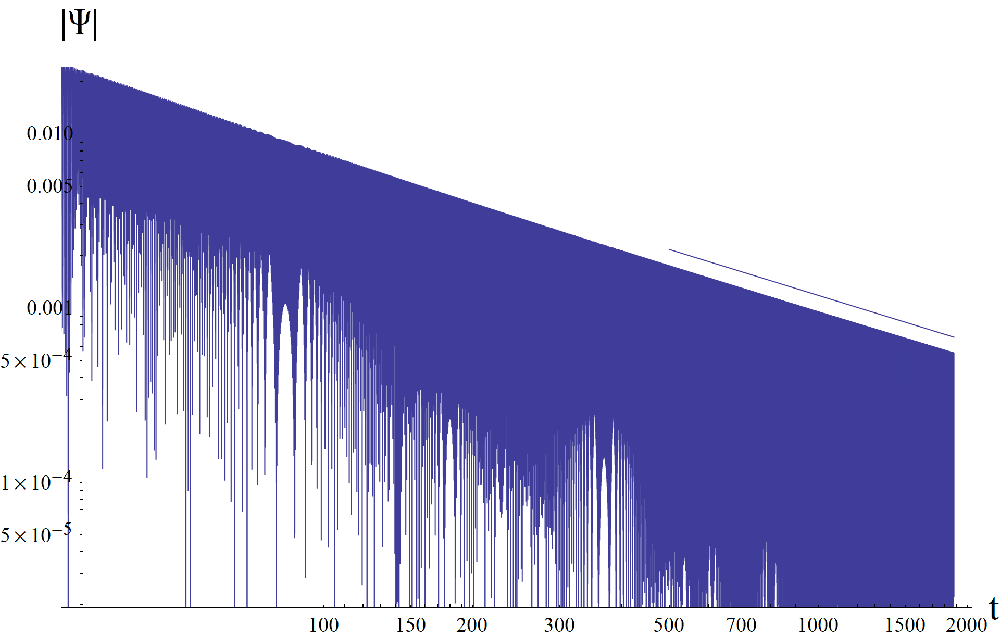}
\caption{Effective potential and logarithmic time-domain profile together with the line $\sim t^{-7/8}$ for $\ell=1$, $\mu=20$, $l_{cr}=1.137$, $M=1$.}\label{fig:L1}
\end{figure*}

Effective potentials for massive fields are shown in Fig.~\ref{fig:gravpot1} for $\ell=0$ and $\ell=1$, respectively. There we can see that at some sufficiently large value of mass $\mu$ the effective potential does not have a peak. The effective potential $V(r)$ remains positive everywhere outside the event horizon. Therefore,  
the operator
\[
\mathcal{D}=-\frac{d^{2}}{dr_*^{2}}+V
\]
\noindent acting on perturbations can be regarded as a positive self-adjoint operator in the Hilbert 
space of square-integrable functions of the tortoise coordinate $r_*$ (see, for instance \cite{Abdalla:2006qj,Konoplya:2007jv}). 
Consequently, any solution of the wave equation with compact support remains bounded, which guarantees stability. Thus, all QNMs must be damped.

\begin{figure*}
\centering
\includegraphics[width=0.48\linewidth]{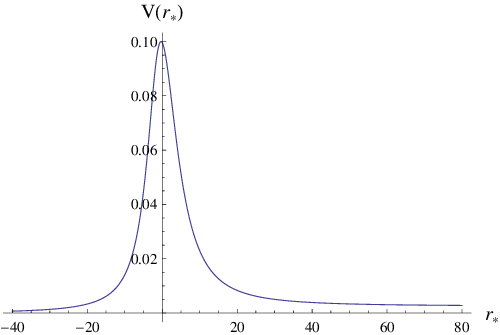}
\includegraphics[width=0.48\linewidth]{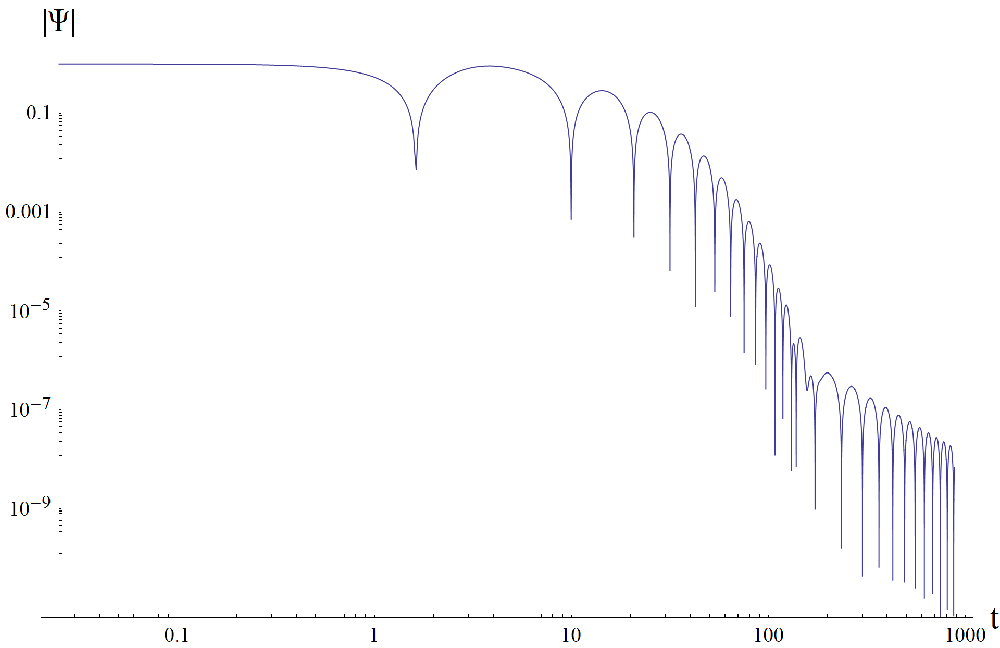}
\caption{Effective potential and logarithmic time-domain profile for $\ell=1$, $\mu=0.05$, $l_{cr}=1.137$, $M=1$. The quasinormal ringing period gives $\omega = 0.288725 - 0.0934358 i$, which is in a good agreement with the 12-th order WKB data $\omega = 0.287974 - 0.093852 i$, while the intermediate tail is close to the law $\sim t^{-5/2}$.}\label{fig:L1mu005}
\end{figure*}

\begin{figure}
\centering
\includegraphics[width=0.99\linewidth]{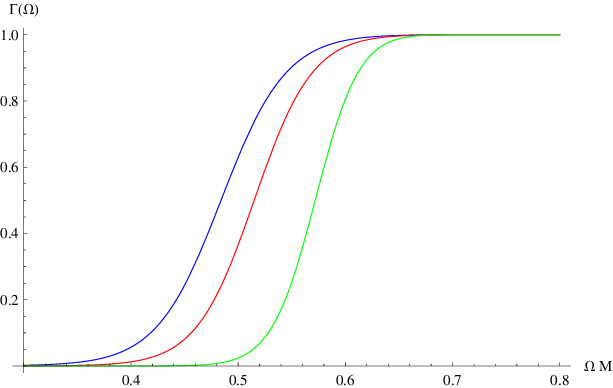}
\caption{Grey-body factors for $\ell=2$, $\mu=0$ (left, blue), $\mu=0.3$ (red, middle) and $\mu=0.5$ (green, right), $l_{cr}=1.137$, $M=1$.}\label{fig:GBF}
\end{figure}

\begin{table}
\begin{tabular}{c c c c}
\hline
\hline
$\mu$ & WKB-10 $m=6$ & WKB-12 $m=7$ & difference \\
\hline
\multicolumn{4}{c}{$\ell=0$} \\
\hline
$0$ & $0.098000-0.087391 i$ & $0.095527-0.098114 i$ & $8.38\%$\\
$0.025$ & $0.099888-0.086984 i$ & $0.096499-0.095897 i$ & $7.20\%$\\
$0.05$ & $0.105662-0.088487 i$ & $0.100180-0.090041 i$ & $4.13\%$\\
$0.075$ & $0.107368-0.083935 i$ & $0.107368-0.083935 i$ & $0.0002\%$\\
$0.1$ & $0.115004-0.080497 i$ & $0.115007-0.080494 i$ & $0.0035\%$\\
$0.125$ & $0.117699-0.078705 i$ & $0.122037-0.077845 i$ & $3.12\%$\\
$0.15$ & $0.112190-0.069169 i$ & $0.111071-0.066540 i$ & $2.17\%$\\
\hline
\multicolumn{4}{c}{$\ell=1$} \\
\hline
$0$ & $0.287007-0.095193 i$ & $0.286621-0.094493 i$ & $0.265\%$\\
$0.05$ & $0.288283-0.094278 i$ & $0.287974-0.093852 i$ & $0.173\%$\\
$0.1$ & $0.292014-0.091613 i$ & $0.292012-0.091613 i$ & $0.0005\%$\\
$0.15$ & $0.298552-0.088438 i$ & $0.297487-0.087829 i$ & $0.394\%$\\
$0.2$ & $0.307787-0.082212 i$ & $0.305741-0.083665 i$ & $0.788\%$\\
$0.25$ & $0.315997-0.076299 i$ & $0.316383-0.077401 i$ & $0.359\%$\\
$0.3$ & $0.331997-0.066147 i$ & $0.329998-0.069558 i$ & $1.17\%$\\
$0.35$ & $0.344611-0.057335 i$ & $0.348020-0.057208 i$ & $0.976\%$\\
$0.4$ & $0.364116-0.053116 i$ & $0.363439-0.052979 i$ & $0.188\%$\\
$0.45$ & $0.388902-0.026066 i$ & $0.389949-0.027174 i$ & $0.391\%$\\
\hline
\hline
\end{tabular}
\caption{Fundamental quasinormal modes ($n=0$) obtained with the WKB method at different orders and Padé approximants for $\ell=0$ and $\ell=1$. The mass is $M=1$ and the critical value of the parameter is $l_{cr}=1.137$.}\label{tab:tableI}
\end{table}

\begin{table}
\begin{tabular}{c c c c}
\hline
\hline
$\mu$ & WKB-10 $m=6$ & WKB-12 $m=7$ & difference \\
\hline
\multicolumn{4}{c}{$\ell=2,\; n=0$} \\
\hline
$0$ & $0.479516-0.095115 i$ & $0.479516-0.095115 i$ & $0.00011\%$\\
$0.05$ & $0.480345-0.094892 i$ & $0.480345-0.094898 i$ & $0.00123\%$\\
$0.1$ & $0.482870-0.094180 i$ & $0.482919-0.094220 i$ & $0.0128\%$\\
$0.15$ & $0.487016-0.092804 i$ & $0.487041-0.092778 i$ & $0.00726\%$\\
$0.2$ & $0.492606-0.090920 i$ & $0.492746-0.090979 i$ & $0.0303\%$\\
$0.25$ & $0.499880-0.088664 i$ & $0.500100-0.088749 i$ & $0.0464\%$\\
$0.3$ & $0.508985-0.085877 i$ & $0.509173-0.085908 i$ & $0.0370\%$\\
$0.35$ & $0.519774-0.082402 i$ & $0.520034-0.082410 i$ & $0.0493\%$\\
$0.4$ & $0.532915-0.077963 i$ & $0.532640-0.078218 i$ & $0.0695\%$\\
$0.45$ & $0.547161-0.072488 i$ & $0.547139-0.073140 i$ & $0.118\%$\\
$0.5$ & $0.562633-0.067449 i$ & $0.563350-0.067077 i$ & $0.143\%$\\
$0.55$ & $0.580664-0.060689 i$ & $0.581000-0.060916 i$ & $0.0694\%$\\
$0.6$ & $0.600838-0.053336 i$ & $0.600645-0.053113 i$ & $0.0489\%$\\
$0.65$ & $0.624905-0.044278 i$ & $0.624880-0.044341 i$ & $0.0109\%$\\
$0.7$ & $0.649139-0.033232 i$ & $0.649117-0.033287 i$ & $0.00913\%$\\
\hline
\multicolumn{4}{c}{$\ell=2,\; n=1$} \\
\hline
$0$ & $0.444805-0.291835 i$ & $0.443861-0.291943 i$ & $0.179\%$\\
$0.05$ & $0.445121-0.291389 i$ & $0.444349-0.291458 i$ & $0.146\%$\\
$0.1$ & $0.446084-0.290077 i$ & $0.445772-0.290080 i$ & $0.0587\%$\\
$0.15$ & $0.447678-0.288036 i$ & $0.447618-0.288026 i$ & $0.0113\%$\\
$0.2$ & $0.449698-0.285734 i$ & $0.461988-0.266926 i$ & $4.22\%$\\
$0.25$ & $0.451638-0.285548 i$ & $0.461811-0.293391 i$ & $2.40\%$\\
$0.3$ & $0.433006-0.285624 i$ & $0.425496-0.286679 i$ & $1.46\%$\\
$0.35$ & $0.445077-0.250901 i$ & $0.445378-0.249812 i$ & $0.221\%$\\
$0.4$ & $0.459056-0.239639 i$ & $0.460199-0.238694 i$ & $0.286\%$\\
$0.45$ & $0.467964-0.228032 i$ & $0.466187-0.229438 i$ & $0.435\%$\\
$0.5$ & $0.480218-0.201980 i$ & $0.472202-0.215237 i$ & $2.97\%$\\
$0.55$ & $0.478415-0.197762 i$ & $0.478490-0.197759 i$ & $0.0145\%$\\
$0.6$ & $0.492658-0.180508 i$ & $0.492177-0.180535 i$ & $0.0919\%$\\
\hline
\multicolumn{4}{c}{$\ell=3,\; n=0$} \\
\hline
$0$ & $0.672308-0.095427 i$ & $0.672333-0.095698 i$ & $0.0402\%$\\
$0.05$ & $0.672912-0.095299 i$ & $0.673145-0.095768 i$ & $0.0771\%$\\
$0.1$ & $0.674726-0.094913 i$ & $0.674906-0.094641 i$ & $0.0478\%$\\
$0.15$ & $0.677758-0.094262 i$ & $0.677787-0.094127 i$ & $0.0201\%$\\
$0.2$ & $0.682009-0.093317 i$ & $0.682000-0.093253 i$ & $0.00933\%$\\
$0.25$ & $0.687415-0.092087 i$ & $0.687427-0.092075 i$ & $0.00245\%$\\
$0.3$ & $0.694073-0.090642 i$ & $0.694121-0.090609 i$ & $0.00831\%$\\
$0.35$ & $0.702001-0.088886 i$ & $0.702044-0.088852 i$ & $0.00787\%$\\
$0.4$ & $0.711175-0.086819 i$ & $0.711223-0.086780 i$ & $0.00862\%$\\
$0.45$ & $0.721610-0.084429 i$ & $0.721676-0.084373 i$ & $0.0119\%$\\
$0.5$ & $0.733363-0.081679 i$ & $0.733419-0.081614 i$ & $0.0117\%$\\
$0.55$ & $0.746489-0.078459 i$ & $0.746473-0.078503 i$ & $0.00623\%$\\
$0.6$ & $0.760868-0.074863 i$ & $0.760867-0.074961 i$ & $0.0129\%$\\
$0.65$ & $0.776333-0.071177 i$ & $0.776619-0.070992 i$ & $0.0436\%$\\
$0.7$ & $0.793776-0.066677 i$ & $0.793767-0.066572 i$ & $0.0132\%$\\
$0.75$ & $0.812200-0.061858 i$ & $0.812412-0.061819 i$ & $0.0265\%$\\
$0.8$ & $0.832432-0.056413 i$ & $0.832355-0.056371 i$ & $0.0105\%$\\
$0.85$ & $0.854290-0.050312 i$ & $0.854326-0.050270 i$ & $0.00640\%$\\
$0.9$ & $0.878227-0.042843 i$ & $0.878077-0.043113 i$ & $0.0351\%$\\
$0.95$ & $0.902213-0.035201 i$ & $0.901985-0.034903 i$ & $0.0417\%$\\
$1.$ & $0.929717-0.026110 i$ & $0.929177-0.026581 i$ & $0.0770\%$\\
\hline
\hline
\end{tabular}
\caption{Quasinormal modes obtained with the WKB method at different orders with Padé approximants for $\ell=2$ ($n=0,1$) and $\ell=3$ ($n=0$); $M=1$, $l_{cr}=1.137$.} \label{tab:tableII}
\end{table}

\section{Quasinormal modes and evolution of perturbations}

The spectrum of quasinormal oscillations can be computed through two complementary approaches: direct numerical evolution of perturbations in the time domain, and semi-analytical expansions based on the WKB formalism. Each has distinct advantages and limitations, especially in the case of massive fields.

\subsection{Time-Domain Integration}
A powerful method for determining quasinormal frequencies is to solve the perturbation equation directly in the time domain. Rewriting the master equation in terms of the null coordinates $u = t-r_*$ and $v = t+r_*$, one discretizes the resulting wave equation on a characteristic grid following the well-known algorithm of~\cite{Gundlach:1993tp}. An initial Gaussian pulse is propagated through the effective potential, producing the full time evolution of the field at a fixed radial position outside the horizon. At intermediate times the signal is dominated by exponentially damped oscillations, i.e.\ the quasinormal ringing, while at late times the waveform transitions into power-law or oscillatory tails depending on the field mass.  

The oscillatory part of the waveform is fitted to a superposition of damped exponentials. This extraction is most efficiently carried out by the Prony method, which provides accurate values for the oscillation frequencies and damping rates. Because it does not rely on the specific shape of the potential, the time-domain approach is especially valuable in situations where semi-analytical approximations lose accuracy, such as for small multipole numbers or for fields with non-negligible mass. 
However, for a massive field with sufficiently large mass, the exponential ringdown is rapidly overtaken by oscillatory tails with a power-law envelope (see Figs.~\ref{fig:L0}–\ref{fig:L1mu005}). As a result, accurate extraction of the quasinormal frequency is possible only for small field masses.
The combined use of time-domain evolution and the Prony fitting technique has been discussed extensively in the literature (see, e.g., \cite{Bolokhov:2023ruj,Lutfuoglu:2025bsf,Konoplya:2020jgt,Malik:2024bmp,Cuyubamba:2016cug,Varghese:2011ku,Lutfuoglu:2025qkt,Dubinsky:2024jqi,Konoplya:2013sba,Qian:2022kaq,Ishihara:2008re,Malik:2025ava,Churilova:2019qph} for details).

\subsection{WKB Approach}
For asymptotically flat black holes the Wentzel–Kramers–Brillouin (WKB) method remains one of the most widely used semi-analytical tools~\cite{Schutz:1985km,Iyer:1986np,Konoplya:2003ii,Matyjasek:2017psv,Konoplya:2019hlu}. Defining
\begin{equation}
Q(r_*) = \omega^{2} - V(r_*), 
\quad V_0 = V(r_{*0}), 
\quad V_0^{(k)} = \left.\frac{d^{k}V}{dr_*^{k}}\right|_{r_{*0}},
\end{equation}
where $r_{*0}$ denotes the location of the potential maximum, one expands $Q(r_*)$ about this point and matches WKB solutions across the turning points. This leads to the $N$-th order quantization condition
\begin{equation}
\frac{i\bigl(\omega^{2}-V_0\bigr)}{\sqrt{-2 V_0^{(2)}}}
-\sum_{k=2}^{N} \Lambda_{k}\!\left(\{V_0^{(j)}\},n\right)
= n+\tfrac{1}{2}, \qquad n=0,1,2,\ldots,
\label{eq:WKB-N}
\end{equation}
with $\Lambda_{k}$ being polynomials in higher derivatives of the potential and in the overtone index $n$.  

Because the WKB series is only asymptotic, its convergence is substantially improved by applying Padé resummation~\cite{Konoplya:2019hlu}. In practice, we employ 6-th and higher WKB order expansions together with balanced Padé approximants (e.g.\ $[3/3]$), which have been demonstrated to deliver excellent accuracy across a variety of settings, ~\cite{Skvortsova:2023zmj,Malik:2023bxc,Bronnikov:2019sbx,Konoplya:2019xmn,Bolokhov:2024ixe,Churilova:2021tgn,Dubinsky:2025fwv,Malik:2024nhy,Skvortsova:2024atk,Dubinsky:2024nzo,Lutfuoglu:2025hjy,Skvortsova:2024wly,Bolokhov:2023bwm,Dubinsky:2024rvf,Dubinsky:2025bvf,Dubinsky:2024hmn,Dubinsky:2024mwd}. For massive fields usage of the 10-12 th orders usually gives the best accuracy, which we observe here via comparison with the time-domain integration,

While the WKB–Padé scheme is highly reliable for massless fields and for moderate values of the effective mass, its applicability becomes restricted in the massive case. When the mass parameter $\mu$ grows sufficiently large, the effective potential no longer possesses the usual single-barrier form and, in certain ranges, the event horizon may even disappear. In such situations the WKB expansion cannot be applied consistently, as the method assumes a potential with two classical turning points. Therefore, for large $\mu$ the time-domain integration remains the primary reliable approach. Conversely, when $\mu$ is small enough that the geometry retains a horizon and a single-barrier potential, WKB calculations provide accurate estimates for the dominant modes, and agreement with time-domain results can be used as a consistency check.

\smallskip
In what follows we employ both approaches: the WKB–Padé method for cases where it is applicable and highly accurate, and the time-domain integration with Prony analysis to validate the results and to explore regions of parameter space where WKB becomes unreliable.

\subsection{Numerical Results for Quasinormal Modes}

From Tables~~\ref{tab:tableI}–\ref{tab:tableII} one can observe a clear trend: as the field mass $\mu$ increases, the real part of the quasinormal frequency steadily grows, while the damping rate decreases significantly. Although the WKB method cannot reliably capture modes with vanishing imaginary part, extrapolation of the data toward larger $\mu$ strongly suggests the existence of nearly undamped modes, commonly referred to as {\it quasi-resonances}. In the time-domain picture, such modes cannot be unambiguously isolated either, since quasinormal oscillations do not form a complete basis and are eventually overtaken by late-time tails. These tails exhibit an oscillatory decay with a power-law envelope, and their remarkably slow falloff can be viewed as a manifestation of the same physical regime that gives rise to quasi-resonances.

In Tables~\ref{tab:tableI}–\ref{tab:tableII} we observe that the quasinormal frequencies vary significantly: the real part changes by several tens of percent, while the damping rate may differ by factors of a few or even by orders of magnitude. At the same time, for $\ell=1$ and higher the expected accuracy of the WKB method is better than one percent. This conclusion follows not only from the consistency of results across different WKB orders and Padé approximants, but also from the direct comparison with time-domain integration (see Fig.~\ref{fig:L1mu005}). Therefore, the relative error of the WKB approach is typically one or more orders of magnitude smaller than the observed effect, providing a reliable criterion for the robustness of our calculations.

Notice that as the multipole number $\ell$ increases, the critical value of the field mass $\mu$ at which the effective potential ceases to possess a maximum also increases. In this regime the WKB method is no longer applicable. Consequently, the maximal value of $\mu$ for which we present the quasinormal frequencies in Tables~~\ref{tab:tableI}–\ref{tab:tableII} differs for each $\ell$.

\subsection{Asymptotic Tails}

At late stages of the evolution, once the quasinormal ringing has sufficiently damped but before the asymptotic regime is reached, the signal enters an {\it intermediate decay phase}. In this regime the perturbing field decreases according to an oscillatory power law,  
\begin{equation}
\Phi(t,r) \sim t^{-p}\,\sin(\mu t+\varphi),
\end{equation}
where $\varphi$ represents a constant phase shift and the decay exponent $p$ is determined both by the multipole index $\ell$ and by the specific properties of the spacetime background. For the Schwarzschild black hole, an observer situated at a fixed radial position measures
\begin{equation}
p=\ell+\tfrac{3}{2},
\end{equation}
a result obtained analytically in a series of pioneering works~\cite{Koyama:2001qw,Koyama:2001ee,Burko:2004jn}, and later extended to some other backgrounds \cite{Lutfuoglu:2025qkt}. Here we observe the same intermediate behavior as for the Schwarzschild case numerically, as illustrated in Fig. \ref{fig:L1mu005}. Notice that the above intermediate asymptotic \cite{Koyama:2000hj,Koyama:2001ee,Koyama:2001qw} is not universal and, for example, for massive fields of other spin the intermediate tails may differ \cite{Konoplya:2006gq}.

At {\it asymptotically late times}, i.e., when $\mu t \gg 1/(\mu^{2} M^{2})$, the dynamics of the massive field changes qualitatively. For  various black-hole backgrounds the decay law transitions into an oscillatory inverse power-law behavior, frequently referred to as the {\it asymptotic tail}, 
\begin{equation}\label{asymptotictail}
\Phi(t,r)\sim t^{-5/6}\,\sin(\mu t+\varphi).
\end{equation}

However, in the case of Dymnikova black hole, we observe a different asymptotic law:
\begin{equation}\label{asymptotictail2}
\Phi(t,r)\sim t^{-7/8}\,\sin(\mu t+\varphi),
\end{equation}
as can be seen for $\ell=0$ and $\ell=1$ in Figs.~\ref{fig:L0} and \ref{fig:L1}.  
This slow falloff, which no longer depends on $\ell$, dominates the ultimate fate of perturbations and reflects the long-range dispersive character of massive fields in black-hole spacetimes.

\vspace{4mm}
\subsection{Relation to Grey-Body Factors}

The QNMs of Dymnikova black holes obtained in this work can be employed to compute the grey-body factors for massive fields through the established correspondence between quasinormal frequencies and transmission coefficients~\cite{Konoplya:2024lir}
\begin{widetext}
\[
\Gamma_{\ell}(\Omega) \approx \left[ 1 + \exp\!\left( \frac{2\pi\bigl(\Omega^{2}-\mathrm{Re}(\omega_{0})^{2}\bigr)}{4\,\mathrm{Re}(\omega_{0})\,\mathrm{Im}(\omega_{0})} \right) \right]^{-1} + \textit{higher order corrections}.
\]
\end{widetext}
Here $\omega_0$ is the fundamental mode and the higher order corrections include the first overtone. Notice that the grey-body factors correspond to boundary conditions which are different from those for  QNMs:
The grey-body factor are
\begin{equation}
\Gamma_{\ell}(\Omega) = 
\left|\frac{A_{\text{trans}}}{A_{\text{in}}}\right|^{2}
= 1 - \left|\frac{A_{\text{out}}}{A_{\text{in}}}\right|^{2},
\end{equation}
where $A_{\text{in}}$, $A_{\text{out}}$, and $A_{\text{trans}}$ are the amplitudes of the incoming, reflected, and transmitted waves, respectively. However, this correspondence holds with high accuracy only in the eikonal limit and becomes merely approximate for low multipoles, as demonstrated in several recent studies~\cite{Lutfuoglu:2025ldc,Malik:2024cgb,Lutfuoglu:2025blw,Bolokhov:2024otn,Han:2025cal,Dubinsky:2025nxv}. Moreover, for massive fields in asymptotically flat spacetimes the applicability of this approach is further restricted: the correspondence remains valid only when the field mass $\mu$ is sufficiently small, since the WKB method underlying the correspondence ceases to provide reliable results at large $\mu$. From Fig.~\ref{fig:GBF} we observe that the mass of the field strongly suppresses the emission of low frequencies via decreasing of the grey-body factors. This could be easily explained from the behavior of the effective potential given in Fig.~\ref{fig:gravpot1}: higher mass $\mu$ corresponds to a higher potential barrier, which means a smaller transmission coefficient.  

\section{Conclusions}  

Quasinormal modes of various regular and quantum-corrected black holes have been extensively investigated in recent years. However, most studies have focused on perturbations of massless fields, while the spectrum of massive fields has received comparatively little attention. In this work, we analyze QNMs of massive scalar perturbations around the Dymnikova regular black hole and demonstrate that their spectrum differs qualitatively from that of massless fields, both in the frequency domain—where long-lived modes emerge—and in the time domain—where distinctive oscillatory tails appear. The grey-body factors are strongly decreased when the mass of the field is turned on, so that emission of massive particles is strongly suppressed in comparison with the massless ones.

It is conceivable that quantum modifications could manifest themselves not only through changes in the underlying geometry but also in more subtle features of particle motion, where certain dynamical aspects may acquire nontrivial corrections~\cite{Hensh:2022xqc,Turimov:2025tmf}. In addition, in settings where matter fields are coupled to the spacetime in ways that go beyond the minimal prescription, or in broader considerations of radiation processes, the specification of boundary conditions may take on a role of unexpected significance~\cite{Cardoso:2019apo,2025EPJB...98...35R,2025JPhA...58H5201D}.

Our analysis has focused on the fundamental mode and, when applicable, the first overtone, for which the WKB approach remains reliable. For massive fields, however, higher overtones are of particular interest: being localized near the horizon \cite{Konoplya:2022pbc,Konoplya:2022hll}, they are likely to exhibit strong sensitivity to quantum corrections, while also interacting nontrivially with the quasi-resonant regime. A rigorous treatment of this sector requires more precise techniques, such as the Frobenius (Leaver) method~\cite{Leaver:1985ax}, combined with suitable rational approximations of the metric. We leave this challenging problem, together with the study of polar perturbations and higher-dimensional extensions, for future work.

From a broader perspective, the Dymnikova geometry serves as a minimal yet robust framework in which to explore how regularity and quantum corrections affect black-hole spectroscopy. Extending the present analysis to other types of perturbed massive fields, to rotating generalizations of the geometry, or to higher-dimensional analogues, could shed further light on the interplay between massive dynamics and near-horizon quantum structure. In this sense, the present work should be viewed as a first step toward a more comprehensive understanding of how massive fields probe regular black holes and, ultimately, the quantum nature of gravity itself.

\vspace{4mm}
\begin{acknowledgments}
BCL is grateful to Excellence Project PrF UHK 2205/2025-2026 for the financial support.
\end{acknowledgments}

\bibliography{bibliography}

@article{Turimov:2025tmf,
    author = "Turimov, Bobur and Usanov, Sulton and Khamroev, Yokubjon",
    title = "{Particles acceleration by Bocharova{\textendash}Bronnikov{\textendash}Melnikov{\textendash}Bekenstein black hole}",
    eprint = "2502.11185",
    archivePrefix = "arXiv",
    primaryClass = "gr-qc",
    doi = "10.1016/j.dark.2025.101876",
    journal = "Phys. Dark Univ.",
    volume = "48",
    pages = "101876",
    year = "2025"
}

@article{Lutfuoglu:2025blw,
    author = {L{\"u}tf{\"u}o{\u{g}}lu, Bekir Can and Saka, Erdin{\c{c}} Ula{\c{s}} and Shermatov, Abubakir and Rayimbaev, Javlon and Ibragimov, Inomjon and Muminov, Sokhibjan},
    title = "{Proper-time approach in asymptotic safety via black hole quasinormal modes and grey-body factors}",
    eprint = "2509.15923",
    archivePrefix = "arXiv",
    primaryClass = "gr-qc",
    doi = "10.1140/epjc/s10052-025-14950-z",
    journal = "Eur. Phys. J. C",
    volume = "85",
    number = "10",
    pages = "1190",
    year = "2025"
}

@ARTICLE{2025EPJB...98...35R,
       author = {{Rakhmanov}, S. and {Matchonov}, K. and {Yusupov}, H. and {Nasriddinov}, K. and {Matrasulov}, D.},
        title = "{Optical high harmonic generation in Dirac materials}",
      journal = {Eur. Phys. J. B},
         year = 2025,
        month = feb,
       volume = {98},
       number = {2},
          eid = {35},
        pages = {35},
          doi = {10.1140/epjb/s10051-025-00885-7},
archivePrefix = {arXiv},
       eprint = {2504.03599},
 primaryClass = {cond-mat.mes-hall},
       adsurl = {https://ui.adsabs.harvard.edu/abs/2025EPJB...98...35R}
}

@ARTICLE{2025JPhA...58H5201D,
       author = {{Derkach}, V.~A. and {Trunk}, C. and {Yusupov}, J.~R. and {Matrasulov}, D.~U.},
        title = "{Transparent boundary conditions for the stationary Schr{\"o}dinger equation via Weyl{\textendash}Titchmarsh theory}",
      journal = {J. Phys. A: Math. Gen.},
         year = 2025,
        month = aug,
       volume = {58},
       number = {34},
          eid = {345201},
        pages = {345201},
          doi = {10.1088/1751-8121/adf787},
archivePrefix = {arXiv},
       eprint = {2410.10232},
 primaryClass = {math-ph},
       adsurl = {https://ui.adsabs.harvard.edu/abs/2025JPhA...58H5201D}
}

@article{LIGOScientific:2016aoc,
    author = "Abbott, B. P. and others",
    collaboration = "LIGO Scientific, Virgo",
    title = "{Observation of Gravitational Waves from a Binary Black Hole Merger}",
    eprint = "1602.03837",
    archivePrefix = "arXiv",
    primaryClass = "gr-qc",
    reportNumber = "LIGO-P150914",
    doi = "10.1103/PhysRevLett.116.061102",
    journal = "Phys. Rev. Lett.",
    volume = "116",
    number = "6",
    pages = "061102",
    year = "2016"
}

@article{LIGOScientific:2017vwq,
    author = "Abbott, B. P. and others",
    collaboration = "LIGO Scientific, Virgo",
    title = "{GW170817: Observation of Gravitational Waves from a Binary Neutron Star Inspiral}",
    eprint = "1710.05832",
    archivePrefix = "arXiv",
    primaryClass = "gr-qc",
    reportNumber = "LIGO-P170817",
    doi = "10.1103/PhysRevLett.119.161101",
    journal = "Phys. Rev. Lett.",
    volume = "119",
    number = "16",
    pages = "161101",
    year = "2017"
}

@article{LIGOScientific:2020zkf,
    author = "Abbott, R. and others",
    collaboration = "LIGO Scientific, Virgo",
    title = "{GW190814: Gravitational Waves from the Coalescence of a 23 Solar Mass Black Hole with a 2.6 Solar Mass Compact Object}",
    eprint = "2006.12611",
    archivePrefix = "arXiv",
    primaryClass = "astro-ph.HE",
    reportNumber = "LIGO-P190814",
    doi = "10.3847/2041-8213/ab960f",
    journal = "Astrophys. J. Lett.",
    volume = "896",
    number = "2",
    pages = "L44",
    year = "2020"
}

@article{Macedo:2024dqb,
    author = "Mac{\^e}do, M. H. and Furtado, J. and Alencar, G. and Landim, R. R.",
    title = "{Thermodynamics and quasinormal modes of the Dymnikova black hole in higher dimensions}",
    eprint = "2404.02818",
    archivePrefix = "arXiv",
    primaryClass = "gr-qc",
    doi = "10.1016/j.aop.2024.169833",
    journal = "Annals Phys.",
    volume = "471",
    pages = "169833",
    year = "2024"
}

@article{Leaver:1985ax,
    author = "Leaver, E. W.",
    title = "{An Analytic representation for the quasi normal modes of Kerr black holes}",
    doi = "10.1098/rspa.1985.0119",
    journal = "Proc. Roy. Soc. Lond. A",
    volume = "402",
    pages = "285--298",
    year = "1985"
}

@article{Bolokhov:2025lnt,
    author = "Bolokhov, S. V. and Skvortsova, Milena",
    title = "{Gravitational Quasinormal Modes and Grey-Body Factors of Bonanno-Reuter Regular Black Holes}",
    eprint = "2507.07196",
    archivePrefix = "arXiv",
    primaryClass = "gr-qc",
    journal = "Int. J. Grav. Theor. Phys.",
    volume = "1",
    pages = "3",
    year = "2025"
}

@article{Bolokhov:2025uxz,
    author = "Bolokhov, Sergei V. and Skvortsova, Milena",
    title = "{Review of Analytic Results on Quasinormal Modes of Black Holes}",
    eprint = "2504.05014",
    archivePrefix = "arXiv",
    primaryClass = "gr-qc",
    doi = "10.1134/S0202289325700306",
    journal = "Grav. Cosmol.",
    volume = "31",
    number = "4",
    pages = "423--446",
    year = "2025"
}

@article{Dymnikova:2003vt,
    author = "Dymnikova, Irina",
    editor = "Mostepanenko, V. M. and Romero, C.",
    title = "{Spherically symmetric space-time with the regular de Sitter center}",
    eprint = "gr-qc/0304110",
    archivePrefix = "arXiv",
    doi = "10.1142/S021827180300358X",
    journal = "Int. J. Mod. Phys. D",
    volume = "12",
    pages = "1015--1034",
    year = "2003"
}

@article{Abdalla:2006qj,
    author = "Abdalla, E. and Cuadros-Melgar, B. and Pavan, A. B. and Molina, C.",
    title = "{Stability and thermodynamics of brane black holes}",
    eprint = "gr-qc/0604033",
    archivePrefix = "arXiv",
    doi = "10.1016/j.nuclphysb.2006.06.017",
    journal = "Nucl. Phys. B",
    volume = "752",
    pages = "40--59",
    year = "2006"
}

@article{Konoplya:2006gq,
    author = "Konoplya, R. A. and Zhidenko, A. and Molina, C.",
    title = "{Late time tails of the massive vector field in a black hole background}",
    eprint = "gr-qc/0602047",
    archivePrefix = "arXiv",
    doi = "10.1103/PhysRevD.75.084004",
    journal = "Phys. Rev. D",
    volume = "75",
    pages = "084004",
    year = "2007"
}

@article{Konoplya:2007jv,
    author = "Konoplya, R. A. and Zhidenko, A.",
    title = "{Stability of multidimensional black holes: Complete numerical analysis}",
    eprint = "hep-th/0703231",
    archivePrefix = "arXiv",
    doi = "10.1016/j.nuclphysb.2007.04.016",
    journal = "Nucl. Phys. B",
    volume = "777",
    pages = "182--202",
    year = "2007"
}

@article{Dymnikova:2004qg,
    author = "Dymnikova, Irina and Galaktionov, Evgeny",
    title = "{Stability of a vacuum nonsingular black hole}",
    eprint = "gr-qc/0409049",
    archivePrefix = "arXiv",
    doi = "10.1088/0264-9381/22/12/003",
    journal = "Class. Quant. Grav.",
    volume = "22",
    pages = "2331--2358",
    year = "2005"
}

@article{KAGRA:2013rdx,
    author = "Abbott, B. P. and others",
    collaboration = "KAGRA, LIGO Scientific, Virgo",
    title = "{Prospects for observing and localizing gravitational-wave transients with Advanced LIGO, Advanced Virgo and KAGRA}",
    eprint = "1304.0670",
    archivePrefix = "arXiv",
    primaryClass = "gr-qc",
    reportNumber = "LIGO-P1200087, VIR-0288A-12, JGW-P1808427",
    doi = "10.1007/s41114-020-00026-9",
    journal = "Living Rev. Rel.",
    volume = "19",
    pages = "1",
    year = "2016"
}

@article{Konoplya:2022pbc,
    author = "Konoplya, R. A. and Zhidenko, A.",
    title = "{First few overtones probe the event horizon geometry}",
    eprint = "2209.00679",
    archivePrefix = "arXiv",
    primaryClass = "gr-qc",
    doi = "10.1016/j.jheap.2024.10.015",
    journal = "JHEAp",
    volume = "44",
    pages = "419--426",
    year = "2024"
}

@article{Konoplya:2022hll,
    author = "Konoplya, R. A. and Zinhailo, A. F. and Kunz, J. and Stuchlik, Z. and Zhidenko, A.",
    title = "{Quasinormal ringing of regular black holes in asymptotically safe gravity: the importance of overtones}",
    eprint = "2206.14714",
    archivePrefix = "arXiv",
    primaryClass = "gr-qc",
    doi = "10.1088/1475-7516/2022/10/091",
    journal = "JCAP",
    volume = "10",
    pages = "091",
    year = "2022"
}

@article{Konoplya:2020jgt,
    author = "Konoplya, R. A. and Zinhailo, A. F. and Stuchlik, Z.",
    title = "{Quasinormal modes and Hawking radiation of black holes in cubic gravity}",
    eprint = "2006.10462",
    archivePrefix = "arXiv",
    primaryClass = "gr-qc",
    doi = "10.1103/PhysRevD.102.044023",
    journal = "Phys. Rev. D",
    volume = "102",
    number = "4",
    pages = "044023",
    year = "2020"
}

@article{Dubinsky:2024nzo,
    author = "Dubinsky, Alexey and Zinhailo, Antonina F.",
    title = "{Analytic expressions for grey-body factors of the general parametrized spherically symmetric black holes}",
    eprint = "2410.15232",
    archivePrefix = "arXiv",
    primaryClass = "gr-qc",
    doi = "10.1209/0295-5075/adbc17",
    journal = "EPL",
    volume = "149",
    number = "6",
    pages = "69004",
    year = "2025"
}

@article{Konoplya:2011qq,
    author = "Konoplya, R. A. and Zhidenko, A.",
    title = "{Quasinormal modes of black holes: From astrophysics to string theory}",
    eprint = "1102.4014",
    archivePrefix = "arXiv",
    primaryClass = "gr-qc",
    doi = "10.1103/RevModPhys.83.793",
    journal = "Rev. Mod. Phys.",
    volume = "83",
    pages = "793--836",
    year = "2011"
}

@article{Berti:2009kk,
    author = "Berti, Emanuele and Cardoso, Vitor and Starinets, Andrei O.",
    title = "{Quasinormal modes of black holes and black branes}",
    eprint = "0905.2975",
    archivePrefix = "arXiv",
    primaryClass = "gr-qc",
    doi = "10.1088/0264-9381/26/16/163001",
    journal = "Class. Quant. Grav.",
    volume = "26",
    pages = "163001",
    year = "2009"
}

@article{Skvortsova:2024msa,
    author = "Skvortsova, Milena",
    title = "{Quantum corrected black holes: testing the correspondence between grey-body factors and quasinormal modes}",
    eprint = "2411.06007",
    archivePrefix = "arXiv",
    primaryClass = "gr-qc",
    doi = "10.1140/epjc/s10052-025-14589-w",
    journal = "Eur. Phys. J. C",
    volume = "85",
    number = "8",
    pages = "854",
    year = "2025"
}

@article{Churilova:2019qph,
    author = "Churilova, M. S. and Konoplya, R. A. and Zhidenko, A.",
    title = "{Arbitrarily long-lived quasinormal modes in a wormhole background}",
    eprint = "1911.05246",
    archivePrefix = "arXiv",
    primaryClass = "gr-qc",
    doi = "10.1016/j.physletb.2020.135207",
    journal = "Phys. Lett. B",
    volume = "802",
    pages = "135207",
    year = "2020"
}

@article{Konoplya:2013sba,
    author = "Konoplya, R. A. and Zhidenko, A.",
    title = "{Instability of D-dimensional extremally charged Reissner-Nordstrom(-de Sitter) black holes: Extrapolation to arbitrary D}",
    eprint = "1309.7667",
    archivePrefix = "arXiv",
    primaryClass = "hep-th",
    doi = "10.1103/PhysRevD.89.024011",
    journal = "Phys. Rev. D",
    volume = "89",
    number = "2",
    pages = "024011",
    year = "2014"
}

@article{Qian:2022kaq,
    author = "Qian, Wei-Liang and Lin, Kai and Shao, Cai-Ying and Wang, Bin and Yue, Rui-Hong",
    title = "{On the late-time tails of massive perturbations in spherically symmetric black holes}",
    eprint = "2203.04477",
    archivePrefix = "arXiv",
    primaryClass = "gr-qc",
    doi = "10.1140/epjc/s10052-022-10910-z",
    journal = "Eur. Phys. J. C",
    volume = "82",
    number = "10",
    pages = "931",
    year = "2022"
}

@article{Varghese:2011ku,
    author = "Varghese, Nijo and Kuriakose, V. C.",
    title = "{Evolution of massive fields around a black hole in Horava gravity}",
    eprint = "1011.6608",
    archivePrefix = "arXiv",
    primaryClass = "gr-qc",
    doi = "10.1007/s10714-011-1201-y",
    journal = "Gen. Rel. Grav.",
    volume = "43",
    pages = "2757--2767",
    year = "2011"
}

@article{Malik:2024bmp,
    author = "Malik, Zainab",
    title = "{Quasinormal modes of the Mannheim{\textendash}Kazanas black holes}",
    doi = "10.1515/zna-2024-0153",
    journal = "Z. Naturforsch. A",
    volume = "79",
    number = "11",
    pages = "1063--1073",
    year = "2024"
}

@article{Dubinsky:2024jqi,
    author = "Dubinsky, Alexey",
    title = "{Telling late-time tails for a massive scalar field in the background of brane-localized black holes}",
    eprint = "2403.01883",
    archivePrefix = "arXiv",
    primaryClass = "gr-qc",
    doi = "10.1209/0295-5075/ad51a3",
    journal = "EPL",
    volume = "147",
    number = "1",
    pages = "19003",
    year = "2024"
}

@article{Cuyubamba:2016cug,
    author = "Cuyubamba, M. A. and Konoplya, R. A. and Zhidenko, A.",
    title = "{Quasinormal modes and a new instability of Einstein-Gauss-Bonnet black holes in the de Sitter world}",
    eprint = "1604.03604",
    archivePrefix = "arXiv",
    primaryClass = "gr-qc",
    doi = "10.1103/PhysRevD.93.104053",
    journal = "Phys. Rev. D",
    volume = "93",
    number = "10",
    pages = "104053",
    year = "2016"
}

@article{Bolokhov:2023ruj,
    author = "Bolokhov, S. V.",
    title = "{Long-lived quasinormal modes and oscillatory tails of the Bardeen spacetime}",
    doi = "10.1103/PhysRevD.109.064017",
    journal = "Phys. Rev. D",
    volume = "109",
    number = "6",
    pages = "064017",
    year = "2024"
}

@article{Chen:2023wkq,
    author = "Chen, H. and Sathiyaraj, T. and Hassanabadi, H. and Yang, Y. and Long, Z. W. and Tu, F. Q.",
    title = "{Quasinormal modes of the EGUP-corrected Schwarzschild black hole}",
    doi = "10.1007/s12648-023-02734-8",
    journal = "Indian J. Phys.",
    volume = "97",
    number = "14",
    pages = "4481--4489",
    year = "2023"
}

@article{Baruah:2023rhd,
    author = {Baruah, Anshuman and {\"O}vg{\"u}n, Ali and Deshamukhya, Atri},
    title = "{Quasinormal modes and bounding greybody factors of GUP-corrected black holes in Kalb{\textendash}Ramond gravity}",
    eprint = "2304.07761",
    archivePrefix = "arXiv",
    primaryClass = "gr-qc",
    doi = "10.1016/j.aop.2023.169393",
    journal = "Annals Phys.",
    volume = "455",
    pages = "169393",
    year = "2023"
}

@article{Moreira:2023cxy,
    author = "Moreira, Zeus S. and Lima Junior, Haroldo C. D. and Crispino, Lu{\'\i}s C. B. and Herdeiro, Carlos A. R.",
    title = "{Quasinormal modes of a holonomy corrected Schwarzschild black hole}",
    eprint = "2302.14722",
    archivePrefix = "arXiv",
    primaryClass = "gr-qc",
    doi = "10.1103/PhysRevD.107.104016",
    journal = "Phys. Rev. D",
    volume = "107",
    number = "10",
    pages = "104016",
    year = "2023"
}

@article{Fu:2023drp,
    author = "Fu, Guoyang and Zhang, Dan and Liu, Peng and Kuang, Xiao-Mei and Wu, Jian-Pin",
    title = "{Peculiar properties in quasinormal spectra from loop quantum gravity effect}",
    eprint = "2301.08421",
    archivePrefix = "arXiv",
    primaryClass = "gr-qc",
    doi = "10.1103/PhysRevD.109.026010",
    journal = "Phys. Rev. D",
    volume = "109",
    number = "2",
    pages = "026010",
    year = "2024"
}

@article{Heidari:2023ssx,
    author = "Heidari, N. and Hassanabadi, H. and Chen, H.",
    title = "{Quantum-corrected scattering of a Schwarzschild black hole with GUP effect}",
    doi = "10.1016/j.physletb.2023.137707",
    journal = "Phys. Lett. B",
    volume = "838",
    pages = "137707",
    year = "2023"
}

@article{Gundlach:1993tp,
    author = "Gundlach, Carsten and Price, Richard H. and Pullin, Jorge",
    title = "{Late time behavior of stellar collapse and explosions: 1. Linearized perturbations}",
    eprint = "gr-qc/9307009",
    archivePrefix = "arXiv",
    reportNumber = "NSF-ITP-93-84",
    doi = "10.1103/PhysRevD.49.883",
    journal = "Phys. Rev. D",
    volume = "49",
    pages = "883--889",
    year = "1994"
}

@article{Lutfuoglu:2025hwh,
    author = {L{\"u}tf{\"u}o{\u{g}}lu, B. C.},
    title = "{Long-lived quasinormal modes around regular black holes and wormholes in Covariant Effective Quantum Gravity}",
    eprint = "2504.09323",
    archivePrefix = "arXiv",
    primaryClass = "gr-qc",
    doi = "10.1088/1475-7516/2025/06/057",
    journal = "JCAP",
    volume = "06",
    pages = "057",
    year = "2025"
}

@article{Konoplya:2020der,
    author = "Konoplya, R. A. and Zhidenko, A.",
    title = "{4D Einstein-Lovelock black holes: Hierarchy of orders in curvature}",
    eprint = "2005.02225",
    archivePrefix = "arXiv",
    primaryClass = "gr-qc",
    doi = "10.1016/j.physletb.2020.135607",
    journal = "Phys. Lett. B",
    volume = "807",
    pages = "135607",
    year = "2020"
}

@article{Bolokhov:2023ozp,
    author = "Bolokhov, Sergey and Bronnikov, Kirill and Konoplya, Roman",
    title = "{Overtones' Outburst and Hawking Evaporation of Kazakov{\textendash}Solodukhin Quantum Corrected Black Hole}",
    eprint = "2306.11083",
    archivePrefix = "arXiv",
    primaryClass = "gr-qc",
    doi = "10.1002/prop.202400187",
    journal = "Fortsch. Phys.",
    volume = "73",
    number = "5",
    pages = "2400187",
    year = "2025"
}

@article{Abdalla:2005hu,
    author = "Abdalla, E. and Konoplya, R. A. and Molina, C.",
    title = "{Scalar field evolution in Gauss-Bonnet black holes}",
    eprint = "hep-th/0507100",
    archivePrefix = "arXiv",
    doi = "10.1103/PhysRevD.72.084006",
    journal = "Phys. Rev. D",
    volume = "72",
    pages = "084006",
    year = "2005"
}

@article{Dubinsky:2025nxv,
    author = "Dubinsky, Alexey",
    title = "{Gravitational perturbations of Dymnikova black holes: Grey-body factors and absorption cross-sections}",
    eprint = "2509.11017",
    archivePrefix = "arXiv",
    primaryClass = "gr-qc",
    doi = "10.1016/j.aop.2025.170299",
    journal = "Annals Phys.",
    volume = "485",
    pages = "170299",
    year = "2026"
}

@article{Ishihara:2008re,
    author = "Ishihara, Hideki and Kimura, Masashi and Konoplya, Roman A. and Murata, Keiju and Soda, Jiro and Zhidenko, Alexander",
    title = "{Evolution of perturbations of squashed Kaluza-Klein black holes: escape from instability}",
    eprint = "0802.0655",
    archivePrefix = "arXiv",
    primaryClass = "hep-th",
    doi = "10.1103/PhysRevD.77.084019",
    journal = "Phys. Rev. D",
    volume = "77",
    pages = "084019",
    year = "2008"
}

@article{Kokkotas:2017zwt,
    author = "Kokkotas, K. and Konoplya, R. A. and Zhidenko, A.",
    title = "{Non-Schwarzschild black-hole metric in four dimensional higher derivative gravity: analytical approximation}",
    eprint = "1705.09875",
    archivePrefix = "arXiv",
    primaryClass = "gr-qc",
    doi = "10.1103/PhysRevD.96.064007",
    journal = "Phys. Rev. D",
    volume = "96",
    pages = "064007",
    year = "2017"
}

@article{Malik:2025ava,
    author = "Malik, Zainab",
    title = {{Long-lived quasinormal modes of brane-localized Reissner{\textendash}Nordstr{\"o}m{\textendash}de Sitter black holes}},
    eprint = "2504.12570",
    archivePrefix = "arXiv",
    primaryClass = "gr-qc",
    doi = "10.1016/j.aop.2025.170238",
    journal = "Annals Phys.",
    volume = "482",
    pages = "170238",
    year = "2025"
}

@article{Konoplya:2023aph,
    author = "Konoplya, R. A. and Stuchlik, Z. and Zhidenko, A. and Zinhailo, A. F.",
    title = "{Quasinormal modes of renormalization group improved Dymnikova regular black holes}",
    eprint = "2303.01987",
    archivePrefix = "arXiv",
    primaryClass = "gr-qc",
    doi = "10.1103/PhysRevD.107.104050",
    journal = "Phys. Rev. D",
    volume = "107",
    number = "10",
    pages = "104050",
    year = "2023"
}

@article{Dubinsky:2024hmn,
    author = "Dubinsky, Alexey and Zinhailo, Antonina",
    title = "{Asymptotic decay and quasinormal frequencies of scalar and Dirac fields around dilaton-de Sitter black holes}",
    eprint = "2404.01834",
    archivePrefix = "arXiv",
    primaryClass = "gr-qc",
    doi = "10.1140/epjc/s10052-024-13206-6",
    journal = "Eur. Phys. J. C",
    volume = "84",
    number = "8",
    pages = "847",
    year = "2024"
}

@article{Dubinsky:2025bvf,
    author = "Dubinsky, Alexey",
    title = "{Long-Lived Quasinormal Modes and Quasi-Resonances around Non-Minimal Einstein-Yang-Mills Black Holes}",
	doi = {10.1140/epjc/s10052-025-14671-3},
    volume = "85",	
    number = {8},
	journal = {Eur. Phys. J. C},
	pages = {924},
    eprint = "2505.08545",
    archivePrefix = "arXiv",
    primaryClass = "gr-qc",
    month = "5",
    year = "2025"
}

@article{Dubinsky:2024rvf,
    author = "Dubinsky, Alexey",
    title = "{Analytic expressions for quasinormal modes of the general parametrized spherically symmetric black holes and the Hod's proposal}",
    eprint = "2409.16569",
    archivePrefix = "arXiv",
    primaryClass = "gr-qc",
    doi = "10.1016/j.physletb.2025.139251",
    journal = "Phys. Lett. B",
    volume = "861",
    pages = "139251",
    year = "2025"
}

@article{Konoplya:2024kih,
    author = "Konoplya, R. A. and Zhidenko, A.",
    title = "{Dymnikova black hole from an infinite tower of higher-curvature corrections}",
    eprint = "2404.09063",
    archivePrefix = "arXiv",
    primaryClass = "gr-qc",
    doi = "10.1016/j.physletb.2024.138945",
    journal = "Phys. Lett. B",
    volume = "856",
    pages = "138945",
    year = "2024"
}

@article{Han:2025cal,
    author = "Han, Hyewon and Gwak, Bogeun",
    title = "{Correspondence between quasinormal modes and grey-body factors in five-dimensional black holes}",
    eprint = "2508.12989",
    archivePrefix = "arXiv",
    primaryClass = "gr-qc",
    month = "8",
    year = "2025"
}

@article{Malik:2025dxn,
    author = "Malik, Zainab",
    title = "{Gravitational Perturbations of the Hayward Spacetime and Testing the Correspondence between Quasinormal Modes and Grey-body Factors}",
    eprint = "2508.19178",
    archivePrefix = "arXiv",
    primaryClass = "gr-qc",
    doi = "10.1007/s10773-025-06198-w",
    journal = "Int. J. Theor. Phys.",
    volume = "64",
    number = "11",
    pages = "314",
    year = "2025"
}

@article{Dymnikova:1992ux,
    author = "Dymnikova, I.",
    title = "{Vacuum nonsingular black hole}",
    doi = "10.1007/BF00760226",
    journal = "Gen. Rel. Grav.",
    volume = "24",
    pages = "235--242",
    year = "1992"
}

@article{Platania:2019kyx,
    author = "Platania, Alessia",
    title = "{Dynamical renormalization of black-hole spacetimes}",
    eprint = "1903.10411",
    archivePrefix = "arXiv",
    primaryClass = "gr-qc",
    doi = "10.1140/epjc/s10052-019-6990-2",
    journal = "Eur. Phys. J. C",
    volume = "79",
    number = "6",
    pages = "470",
    year = "2019"
}

@article{Bolokhov:2024ixe,
    author = "Bolokhov, S. V.",
    title = "{Late time decay of scalar and Dirac fields around an asymptotically de Sitter black hole in the Euler{\textendash}Heisenberg electrodynamics}",
    eprint = "2404.09364",
    archivePrefix = "arXiv",
    primaryClass = "gr-qc",
    doi = "10.1140/epjc/s10052-024-12990-5",
    journal = "Eur. Phys. J. C",
    volume = "84",
    number = "6",
    pages = "634",
    year = "2024"
}

@article{Iyer:1986np,
    author = "Iyer, Sai and Will, Clifford M.",
    title = "{Black Hole Normal Modes: A {WKB} Approach. 1. Foundations and Application of a Higher Order {WKB} Analysis of Potential Barrier Scattering}",
    reportNumber = "Print-86-1482 (WASH. U., ST. LOUIS)",
    doi = "10.1103/PhysRevD.35.3621",
    journal = "Phys. Rev. D",
    volume = "35",
    pages = "3621",
    year = "1987"
}

@article{Konoplya:2003ii,
    author = "Konoplya, R. A.",
    title = "{Quasinormal behavior of the d-dimensional Schwarzschild black hole and higher order WKB approach}",
    eprint = "gr-qc/0303052",
    archivePrefix = "arXiv",
    doi = "10.1103/PhysRevD.68.024018",
    journal = "Phys. Rev. D",
    volume = "68",
    pages = "024018",
    year = "2003"
}

@article{Konoplya:2019hlu,
    author = "Konoplya, R. A. and Zhidenko, A. and Zinhailo, A. F.",
    title = "{Higher order WKB formula for quasinormal modes and grey-body factors: recipes for quick and accurate calculations}",
    eprint = "1904.10333",
    archivePrefix = "arXiv",
    primaryClass = "gr-qc",
    doi = "10.1088/1361-6382/ab2e25",
    journal = "Class. Quant. Grav.",
    volume = "36",
    pages = "155002",
    year = "2019"
}

@article{Skvortsova:2024wly,
    author = "Skvortsova, Milena",
    title = "{Ringing of Extreme Regular Black Holes}",
    eprint = "2405.15807",
    archivePrefix = "arXiv",
    primaryClass = "gr-qc",
    doi = "10.1134/S020228932470018X",
    journal = "Grav. Cosmol.",
    volume = "30",
    number = "3",
    pages = "279--288",
    year = "2024"
}

@article{Skvortsova:2024atk,
    author = "Skvortsova, Milena",
    title = "{Quasinormal Frequencies of Fields with Various Spin in the Quantum Oppenheimer{\textendash}Snyder Model of Black Holes}",
    eprint = "2405.06390",
    archivePrefix = "arXiv",
    primaryClass = "gr-qc",
    doi = "10.1002/prop.202400132",
    journal = "Fortsch. Phys.",
    volume = "72",
    number = "9-10",
    pages = "2400132",
    year = "2024"
}

@article{Bolokhov:2023bwm,
    author = "Bolokhov, S. V.",
    title = "{Long-lived quasinormal modes and overtones{\textquoteright} behavior of holonomy-corrected black holes}",
    eprint = "2311.05503",
    archivePrefix = "arXiv",
    primaryClass = "gr-qc",
    doi = "10.1103/PhysRevD.110.024010",
    journal = "Phys. Rev. D",
    volume = "110",
    number = "2",
    pages = "024010",
    year = "2024"
}

@article{Dubinsky:2025fwv,
    author = "Dubinsky, Alexey",
    title = "{Black Holes Immersed in Galactic Dark Matter Halo}",
    eprint = "2507.00256",
    archivePrefix = "arXiv",
    primaryClass = "gr-qc",
    journal = "Int. J. Grav. Theor. Phys.",
    volume = "1",
    pages = "2",
    year = "2025"
}

@article{Kokkotas:1999bd,
    author = "Kokkotas, Kostas D. and Schmidt, Bernd G.",
    title = "{Quasinormal modes of stars and black holes}",
    eprint = "gr-qc/9909058",
    archivePrefix = "arXiv",
    doi = "10.12942/lrr-1999-2",
    journal = "Living Rev. Rel.",
    volume = "2",
    pages = "2",
    year = "1999"
}

@article{Lutfuoglu:2025ldc,
    author = {L{\"u}tf{\"u}o{\u{g}}lu, Bekir Can},
    title = "{Black Holes in Proca-Gauss-Bonnet Gravity with Primary Hair: Particle Motion, Shadows, and Grey-Body Factors}",
    eprint = "2507.09246",
    archivePrefix = "arXiv",
    primaryClass = "gr-qc",
    journal = "Int.  J.  Grav.  Theor.  Phys",
    volume = "1",
    pages = "4",
    year = "2025"
}

@article{Malik:2024cgb,
    author = "Malik, Zainab",
    title = "{Correspondence between quasinormal modes and grey-body factors for massive fields in Schwarzschild-de~Sitter spacetime}",
    eprint = "2412.19443",
    archivePrefix = "arXiv",
    primaryClass = "gr-qc",
    doi = "10.1088/1475-7516/2025/04/042",
    journal = "JCAP",
    volume = "04",
    pages = "042",
    year = "2025"
}

@article{Bonanno:2025dry,
    author = "Bonanno, Alfio M. and Konoplya, Roman A. and Oglialoro, Giovanni and Spina, Andrea",
    title = "{Regular black holes from proper-time flow in quantum gravity and their quasinormal modes, shadow and Hawking radiation}",
    eprint = "2509.12469",
    archivePrefix = "arXiv",
    primaryClass = "gr-qc",
    doi = "10.1088/1475-7516/2025/12/042",
    journal = "JCAP",
    volume = "12",
    pages = "042",
    year = "2025"
}

@article{Lutfuoglu:2025bsf,
    author = {L{\"u}tf{\"u}o{\u{g}}lu, B. C.},
    title = "{Long-lived quasinormal modes in the Euler-Heisenberg electrodynamics}",
    eprint = "2508.13361",
    archivePrefix = "arXiv",
    primaryClass = "gr-qc",
    doi = "10.1016/j.physletb.2025.140026",
    journal = "Phys. Lett. B",
    volume = "871",
    pages = "140026",
    year = "2025"
}

@article{Lutfuoglu:2025hjy,
    author = {L{\"u}tf{\"u}o{\u{g}}lu, B. C.},
    title = "{Long-lived quasinormal modes and gray-body factors of black holes and wormholes in dark matter inspired Weyl gravity}",
    eprint = "2503.16087",
    archivePrefix = "arXiv",
    primaryClass = "gr-qc",
    doi = "10.1140/epjc/s10052-025-14210-0",
    journal = "Eur. Phys. J. C",
    volume = "85",
    number = "5",
    pages = "486",
    year = "2025"
}

@article{Konoplya:2019xmn,
    author = "Konoplya, R. A.",
    title = "{Quantum corrected black holes: quasinormal modes, scattering, shadows}",
    eprint = "1912.10582",
    archivePrefix = "arXiv",
    primaryClass = "gr-qc",
    doi = "10.1016/j.physletb.2020.135363",
    journal = "Phys. Lett. B",
    volume = "804",
    pages = "135363",
    year = "2020"
}

@article{Konoplya:2024lir,
    author = "Konoplya, R. A. and Zhidenko, A.",
    title = "{Correspondence between grey-body factors and quasinormal modes}",
    eprint = "2406.11694",
    archivePrefix = "arXiv",
    primaryClass = "gr-qc",
    doi = "10.1088/1475-7516/2024/09/068",
    journal = "JCAP",
    volume = "09",
    pages = "068",
    year = "2024"
}

@article{Konoplya:2004wg,
    author = "Konoplya, R. A. and Zhidenko, A. V.",
    title = "{Decay of massive scalar field in a Schwarzschild background}",
    eprint = "gr-qc/0411059",
    archivePrefix = "arXiv",
    doi = "10.1016/j.physletb.2005.01.078",
    journal = "Phys. Lett. B",
    volume = "609",
    pages = "377--384",
    year = "2005"
}

@article{Aragon:2020teq,
    author = "Arag{\'o}n, Almendra and B{\'e}car, Ram{\'o}n and Gonz{\'a}lez, P. A. and V{\'a}squez, Yerko",
    title = "{Massive Dirac quasinormal modes in Schwarzschild{\textendash}de Sitter black holes: Anomalous decay rate and fine structure}",
    eprint = "2009.09436",
    archivePrefix = "arXiv",
    primaryClass = "gr-qc",
    doi = "10.1103/PhysRevD.103.064006",
    journal = "Phys. Rev. D",
    volume = "103",
    number = "6",
    pages = "064006",
    year = "2021"
}

@article{Ponglertsakul:2020ufm,
    author = "Ponglertsakul, Supakchai and Gwak, Bogeun",
    title = "{Massive scalar perturbations on Myers-Perry{\textendash}de Sitter black holes with a single rotation}",
    eprint = "2007.16108",
    archivePrefix = "arXiv",
    primaryClass = "gr-qc",
    doi = "10.1140/epjc/s10052-020-08616-1",
    journal = "Eur. Phys. J. C",
    volume = "80",
    number = "11",
    pages = "1023",
    year = "2020"
}

@article{Gonzalez:2022upu,
    author = "Gonz{\'a}lez, P. A. and Papantonopoulos, Eleftherios and Saavedra, Joel and V{\'a}squez, Yerko",
    title = {{Quasinormal modes for massive charged scalar fields in Reissner-Nordstr{\"o}m dS black holes: anomalous decay rate}},
    eprint = "2204.01570",
    archivePrefix = "arXiv",
    primaryClass = "gr-qc",
    doi = "10.1007/JHEP06(2022)150",
    journal = "JHEP",
    volume = "06",
    pages = "150",
    year = "2022"
}

@article{Burikham:2017gdm,
    author = "Burikham, Piyabut and Ponglertsakul, Supakchai and Tannukij, Lunchakorn",
    title = "{Charged scalar perturbations on charged black holes in de Rham-Gabadadze-Tolley massive gravity}",
    eprint = "1709.02716",
    archivePrefix = "arXiv",
    primaryClass = "gr-qc",
    doi = "10.1103/PhysRevD.96.124001",
    journal = "Phys. Rev. D",
    volume = "96",
    number = "12",
    pages = "124001",
    year = "2017"
}

@article{Seahra:2004fg,
    author = "Seahra, Sanjeev S. and Clarkson, Chris and Maartens, Roy",
    title = "{Detecting extra dimensions with gravity wave spectroscopy: the black string brane-world}",
    eprint = "gr-qc/0408032",
    archivePrefix = "arXiv",
    doi = "10.1103/PhysRevLett.94.121302",
    journal = "Phys. Rev. Lett.",
    volume = "94",
    pages = "121302",
    year = "2005"
}

@article{Churilova:2020bql,
    author = "Churilova, M. S.",
    title = "{Black holes in Einstein-aether theory: Quasinormal modes and time-domain evolution}",
    eprint = "2002.03450",
    archivePrefix = "arXiv",
    primaryClass = "gr-qc",
    doi = "10.1103/PhysRevD.102.024076",
    journal = "Phys. Rev. D",
    volume = "102",
    number = "2",
    pages = "024076",
    year = "2020"
}

@article{Zinhailo:2024jzt,
    author = "Zinhailo, Antonina F.",
    title = "{Exploring unique quasinormal modes of a massive scalar field in brane-world scenarios}",
    eprint = "2403.06867",
    archivePrefix = "arXiv",
    primaryClass = "gr-qc",
    doi = "10.1016/j.physletb.2024.138682",
    journal = "Phys. Lett. B",
    volume = "853",
    pages = "138682",
    year = "2024"
}

@article{Jing:2004zb,
    author = "Jing, Jiliang",
    title = "{Late-time evolution of charged massive Dirac fields in the Reissner-Nordstrom black-hole background}",
    eprint = "gr-qc/0408090",
    archivePrefix = "arXiv",
    doi = "10.1103/PhysRevD.72.027501",
    journal = "Phys. Rev. D",
    volume = "72",
    pages = "027501",
    year = "2005"
}

@article{Koyama:2001qw,
    author = "Koyama, Hiroko and Tomimatsu, Akira",
    title = "{Slowly decaying tails of massive scalar fields in spherically symmetric space-times}",
    eprint = "gr-qc/0112075",
    archivePrefix = "arXiv",
    doi = "10.1103/PhysRevD.65.084031",
    journal = "Phys. Rev. D",
    volume = "65",
    pages = "084031",
    year = "2002"
}

@article{Moderski:2001tk,
    author = "Moderski, Rafal and Rogatko, Marek",
    title = "{Late time evolution of a selfinteracting scalar field in the space-time of dilaton black hole}",
    eprint = "gr-qc/0105056",
    archivePrefix = "arXiv",
    doi = "10.1103/PhysRevD.64.044024",
    journal = "Phys. Rev. D",
    volume = "64",
    pages = "044024",
    year = "2001"
}

@article{Rogatko:2007zz,
    author = "Rogatko, Marek and Szyplowska, Agnieszka",
    title = "{Decay of massive scalar hair on brane black holes}",
    doi = "10.1103/PhysRevD.76.044010",
    journal = "Phys. Rev. D",
    volume = "76",
    pages = "044010",
    year = "2007"
}

@article{Koyama:2001ee,
    author = "Koyama, Hiroko and Tomimatsu, Akira",
    title = "{Asymptotic tails of massive scalar fields in Schwarzschild background}",
    eprint = "gr-qc/0103086",
    archivePrefix = "arXiv",
    doi = "10.1103/PhysRevD.64.044014",
    journal = "Phys. Rev. D",
    volume = "64",
    pages = "044014",
    year = "2001"
}

@article{Koyama:2000hj,
    author = "Koyama, Hiroko and Tomimatsu, Akira",
    title = "{Asymptotic power law tails of massive scalar fields in Reissner-Nordstrom background}",
    eprint = "gr-qc/0012022",
    archivePrefix = "arXiv",
    doi = "10.1103/PhysRevD.63.064032",
    journal = "Phys. Rev. D",
    volume = "63",
    pages = "064032",
    year = "2001"
}

@article{Gibbons:2008gg,
    author = "Gibbons, Gary W. and Rogatko, Marek and Szyplowska, Agnieszka",
    title = "{Decay of Massive Dirac Hair on a Brane-World Black Hole}",
    eprint = "0802.3259",
    archivePrefix = "arXiv",
    primaryClass = "hep-th",
    doi = "10.1103/PhysRevD.77.064024",
    journal = "Phys. Rev. D",
    volume = "77",
    pages = "064024",
    year = "2008"
}

@article{Gibbons:2008rs,
    author = "Gibbons, Gary W. and Rogatko, Marek",
    title = "{The Decay of Dirac Hair around a Dilaton Black Hole}",
    eprint = "0801.3130",
    archivePrefix = "arXiv",
    primaryClass = "hep-th",
    doi = "10.1103/PhysRevD.77.044034",
    journal = "Phys. Rev. D",
    volume = "77",
    pages = "044034",
    year = "2008"
}

@article{Konoplya:2008hj,
    author = "Konoplya, R. A.",
    title = "{Magnetic field creates strong superradiant instability}",
    eprint = "0801.0846",
    archivePrefix = "arXiv",
    primaryClass = "hep-th",
    doi = "10.1016/j.physletb.2008.11.059",
    journal = "Phys. Lett. B",
    volume = "666",
    pages = "283--287",
    year = "2008"
}

@article{Wu:2015fwa,
    author = "Wu, Chen and Xu, Renli",
    title = "{Decay of massive scalar field in a black hole background immersed in magnetic field}",
    eprint = "1507.04911",
    archivePrefix = "arXiv",
    primaryClass = "gr-qc",
    doi = "10.1140/epjc/s10052-015-3632-1",
    journal = "Eur. Phys. J. C",
    volume = "75",
    number = "8",
    pages = "391",
    year = "2015"
}

@article{Konoplya:2023fmh,
    author = "Konoplya, R. A. and Zhidenko, A.",
    title = "{Asymptotic tails of massive gravitons in light of pulsar timing array observations}",
    eprint = "2307.01110",
    archivePrefix = "arXiv",
    primaryClass = "gr-qc",
    doi = "10.1016/j.physletb.2024.138685",
    journal = "Phys. Lett. B",
    volume = "853",
    pages = "138685",
    year = "2024"
}

@article{NANOGrav:2023hvm,
    author = "Afzal, Adeela and others",
    collaboration = "NANOGrav",
    title = "{The NANOGrav 15 yr Data Set: Search for Signals from New Physics}",
    eprint = "2306.16219",
    archivePrefix = "arXiv",
    primaryClass = "astro-ph.HE",
    reportNumber = "FERMILAB-PUB-23-589-T",
    doi = "10.3847/2041-8213/acdc91",
    journal = "Astrophys. J. Lett.",
    volume = "951",
    number = "1",
    pages = "L11",
    year = "2023",
    note = "[Erratum: Astrophys.J.Lett. 971, L27 (2024), Erratum: Astrophys.J. 971, L27 (2024)]"
}

@article{Fernandes:2021qvr,
    author = "Fernandes, Tiago V. and Hilditch, David and Lemos, Jos{\'e} P. S. and Cardoso, Vitor",
    title = "{Quasinormal modes of Proca fields in a Schwarzschild-AdS spacetime}",
    eprint = "2112.03282",
    archivePrefix = "arXiv",
    primaryClass = "gr-qc",
    doi = "10.1103/PhysRevD.105.044017",
    journal = "Phys. Rev. D",
    volume = "105",
    number = "4",
    pages = "044017",
    year = "2022"
}

\end{document}